\documentclass{article}

\usepackage{microtype}
\usepackage{graphicx}
\usepackage{subfigure}
\usepackage{makecell}
\usepackage{multirow}
\usepackage{enumerate}
\usepackage{tabularx}
\usepackage{amsmath,amsfonts,amsthm}
\usepackage{booktabs} 

\def\0{{\bf 0}}
\def\1{{\bf 1}}

\def\ie{{\em i.e.}}

\usepackage{hyperref}

\usepackage[accepted]{icml2023}

\usepackage{amssymb}
\usepackage{mathtools}

\usepackage[capitalize,noabbrev]{cleveref}

\theoremstyle{plain}

\theoremstyle{definition}

\theoremstyle{remark}

\usepackage[textsize=tiny]{todonotes}

\icmltitlerunning{TBDetector:Transformer-Based Detector for Advanced Persistent Threats with Provenance Graph}

\begin{document}

\twocolumn[
\icmltitle{TBDetector:Transformer-Based Detector for Advanced Persistent Threats \\ with Provenance Graph}



\icmlsetsymbol{equal}{*}

\begin{icmlauthorlist}
  \icmlauthor{Nan Wang}{}
  \icmlauthor{Xuezhi Wen*}{}
  \icmlauthor{Dalin	Zhang}{}
  \icmlauthor{Xibin	Zhao}{}
  \icmlauthor{Jiahui Ma}{}
  \icmlauthor{Mengxia Luo}{}
  \icmlauthor{Fan Xu}{}
  \icmlauthor{Sen Nie}{}
  \icmlauthor{Shi Wu}{}
  \icmlauthor{Jiqiang Liu}{}
\end{icmlauthorlist}

\icmlkeywords{Machine Learning, ICML}

\vskip 0.3in
]




\begin{abstract}
APT detection is difficult to detect due to the long-term latency, covert and slow multistage attack patterns of Advanced Persistent Threat (APT). 
To tackle these issues, we propose TBDetector, a transformer-based advanced persistent threat detection method for APT attack detection. 
Considering that provenance graphs provide rich historical information and have the powerful attacks historic correlation ability to identify anomalous activities, 
TBDetector employs provenance analysis for APT detection, which summarizes long-running system execution with space efficiency and utilizes transformer with self-attention based encoder-decoder to extract long-term contextual features of system states to detect slow-acting attacks. 
Furthermore, we further introduce anomaly scores to investigate the anomaly of different system states, where each state is calculated with an anomaly score corresponding to its similarity score and isolation score. 
To evaluate the effectiveness of the proposed method, we have conducted experiments on five public datasets, \ie, streamspot, cadets, shellshock, clearscope, and wget\_baseline. 
Experimental results and comparisons with state-of-the-art methods have exhibited better performance of our proposed method. 
\vspace{-0.2cm}
\end{abstract}

\section{Introduction}

Advanced persistent threat attacks are becoming increasingly widespread as an efficient and precise style of cyber attack, 
and they are swiftly becoming one of the most significant risks to enterprise information security. 
APT attack patterns are "low-and-slow" and frequent use of zero-day exploits. 
Compared with traditional network attacks, 
it has five significant characteristics: strong targeting, well-organized, long duration, high concealment, and indirect attack. 

\begin{figure}[h]
  \centering
  \includegraphics[width=240pt]{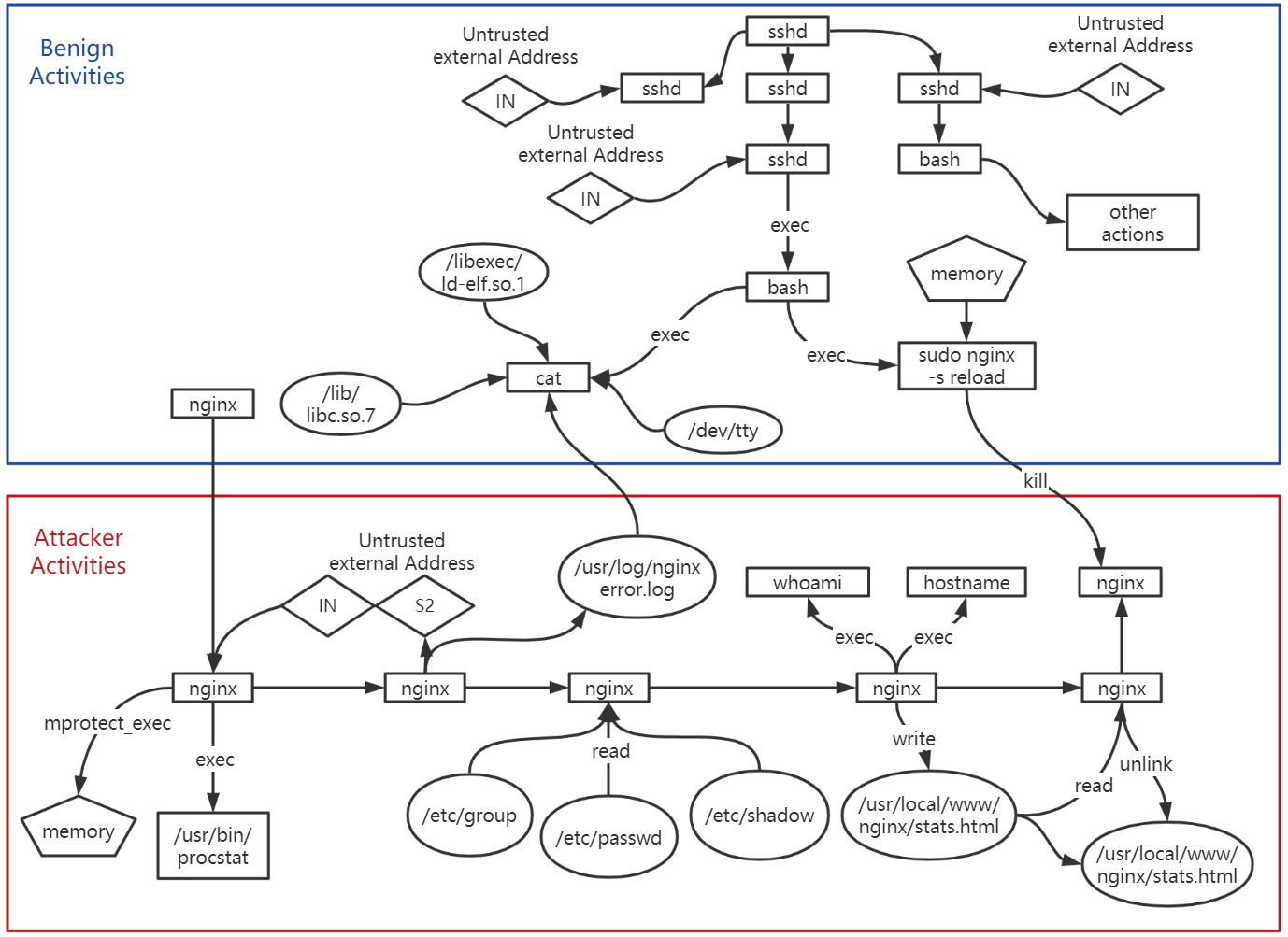}
  \caption{Provenance graph of the running example.}
  \label{figure:Provenance Graph of the Running Example}
  \vspace{-0.7cm}
\end{figure}

Due to the aforementioned APT attack characteristics, the behaviors of APT attacks must be recorded in the APT attack detection process. 
Since provenance graphs can record contextual information about APT behavior, 
more and more research is focusing on using them to record APT attack behavior\cite{Unicorn} \cite{seqnet} \cite{NoDoze} \cite{PASS} \cite{R101}. 
In recent years, the provenance graph has emerged as a promising research direction in the field of intrusion detection. 
An example of a provenance graph is demonstrated in Figure \ref{figure:Provenance Graph of the Running Example}. 
The nodes in the graph represent system entities like processes, files, and memory objects. 
Edges represent system calls and point in the direction of causality. 
As a result, the provenance graph can accurately model the interaction between system entities and restore various system behaviors. 

The APT detection research based on the provenance graph can be mainly divided into three directions: 
graph matching-based detection\cite{de2008subgraph} \cite{POIROT} \cite{Unicorn} \cite{goyal2018graph} \cite{wang2014graph} \cite{yan2006graph} \cite{Log2vec} \cite{yawyd}, 
anomaly score-based detection\cite{Pagoda} \cite{P-Gaussian} \cite{PrioTracker} \cite{NoDoze}, 
and tag propagation-based detection\cite{jiang2006} \cite{SLEUTH} \cite{HOLMES} \cite{MORSE}. 
Most of this anomaly score-based and tag propagation-based attack detection methods are difficult to model long-term behavioral patterns. 
The detection method based on graph matching leads to the continuous growth of the graph size with APT penetration, 
which makes the calculation heavy and may lead to a large amount of information loss. 
Despite the fact that these efforts focus on APT detection issues, 
the contextual analysis of provenance graph analysis and the required long-term characteristics exploring APT attacks are computationally demanding. 
As APT slowly permeates the system, the size of the graph grows. 
In addition, previously unknown APT attacks, \ie, zero-day threats, which may cause serious security threats and financial losses, making the mission more challenging. 

To address the above issues, we propose a new method named TBDetector. 
The method is mainly divided into three stages: 
the first stage converts the provenance graph into a characteristic sequence that reflects system changes. 
The second stage uses the encoder-decoder of the self-attention mechanism to train a sequence feature extractor to extract the long-term contextual features of the feature sequence. 
The third stage uses the similarity score and the isolation score to calculate the anomaly score for each example to detect APT attacks. 
To evaluate the effectiveness of the proposed method, we have conducted experiments on five public datasets, 
\ie, streamspot, cadets, shellshock, clearscope, and wget\_baseline. 
Experimental results and comparisons with state-of-the-art methods have exhibited better performance of our proposed method. 
The main contributions of this paper are as follows: 
\vspace{-0.3cm}
\begin{enumerate}[(1)]
\item TBDetector generates the log information into a provenance graph and converts it into long sequence features. 
Then the encoder-decoder model of self-attention mechanism is used to train a sequence feature extractor, 
which can effectively extract long-term contextual features. Long-term contextual features that have been obtained are able to record APT behavior, 
making it simpler for the detector to tell normal data from anomaly data. 
\item In order to detect the unknown APT attacks, we calculate the anomaly score of each piece of data, 
which is consist of a similarity score and an isolation score. 
\item Five open-source datasets are used to test the proposed approach \ie, streamspot, cadets, shellshock, clearscope, and wget\_baseline. 
Experimental results show that our proposed method has a better performance compared to the state-of-the-art methods. 
\end{enumerate}

\vspace{-0.3cm}
\section{Related work}
\begin{figure*}
  \centering
  \includegraphics[width=17cm]{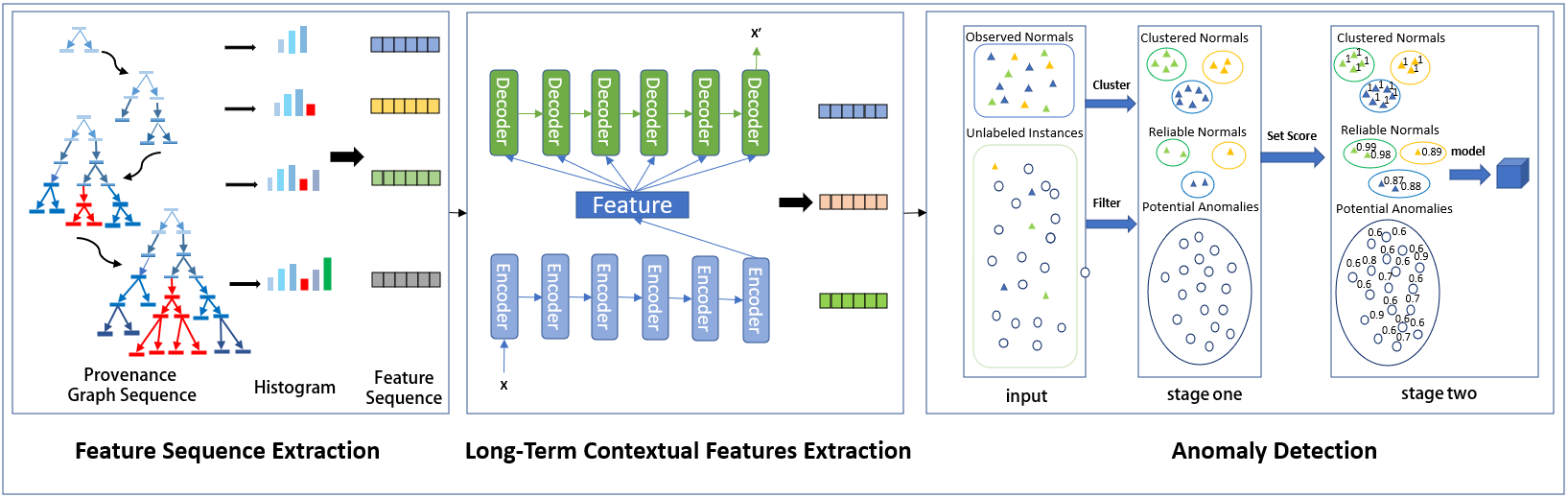}
  \caption{TBDetector frame diagram}
  \label{figure:TBDetector frame diagram}
  \vspace{-0.5cm}
\end{figure*}
Provenance graphs have been used extensively in APT detection in recent years due to the superiority of connecting nodes with causal relationships and representing data flow and control flow relationships between system objects. 
Both academics and industry are paying increasing attention to this type of learning method\cite{Unicorn} \cite{seqnet} \cite{NoDoze} \cite{PASS}. 
APT detection based on provenance graph is mainly divided into three directions: graph matching-based detection, anomaly score-based detection, and tag propagation-based detection. 

The first is graph matching-based detection. 
Due to the substructure in the provenance graph can completely describe malicious behavior, 
it is a very popular method to use graph matching-based detection method to detect APT attacks. 
For example, 
Poirot\cite{POIROT} threat detection is modeled as an imprecise graph pattern matching problem. 
A graph matching method is proposed to identify attacks in provenance graphs. 
But it needs to construct the attack graph according to the prior knowledge, and cannot detect the unknown attack method. 
UNICORN\cite{Unicorn} is the first APT intrusion detection system to analyze the local complete system operation. 
The provenance graph is converted into a sequence of feature vectors and uses an automaton to model the clustering. 
However, the effectiveness of detection is negatively impacted by a large number of automatic opportunities, and the ability to describe the feature sequence is weak. 
In order to preserve as much of the structure and information of graphics as possible, 
graphic embedding is widely used to extract graphic features into vertices\cite{goyal2018graph} \cite{wang2014graph} \cite{yan2006graph}. 
Log2vec\cite{Log2vec} and ProvDetector\cite{yawyd} separate malicious and benign log entries into different clusters, 
recognize malicious logs and effectively detect threats. 
However, there are some drawbacks as well. 
The construction of a query graph requires more prior knowledge. 
It is difficult to detect unknown APT attacks through simple matching rules based on existing attack knowledge. 

The second is anomaly score-based detection. 
The anomaly connections in the system can be found through the statistics of the historical data of each edge. 
The outlier-based detection method is simpler to implement and requires fewer parameters to adjust than the graph-based anomaly detection model. 
For instance, 
Pagoda\cite{Pagoda} takes into account the degree of anomaly for both the entire provenance graph and a specific provenance path. 
P-Gaussian\cite{P-Gaussian} subsequence work can use a gaussian distribution scheme to detect variants. 
PrioTracker\cite{PrioTracker} and NoDoze\cite{NoDoze} adjust the events' suspiciousness based on their neighbor's suspiciousness. 

The third is tag propagation-based detection. 
This part can be divided into two stages: tag initialization and tag propagation. 
Since the number of nodes is much smaller than the number of edges, it is efficient to store and update labels. 
For example, 
\cite{jiang2006} propose a label based simplification process coloring method. 
In the initialization phase, tags are assigned to each remotely accessible server or process. 
During the tag propagation phase, tags can be inherited through derived sub processes, or indirectly spread through process operations. 
SLEUTH\cite{SLEUTH} introduced the provenance graph for the first time to detect APT attacks, used causal relationship tracking and provenance graph to construct a model, and proposed a method and system that can reconstruct attack scenarios in real time on enterprise hosts. By defining rules to match higher-threat attacks, the corresponding scores will be generated. The scene graph will be generated and restored after the tag-based analysis. 
Holmes\cite{HOLMES} integrated HSG and ATT\&CK framework to solve the problem of semantic loss added denoising and pruning and other operations, constructed a system that could detect APT attacks in real-time, and mapped APT activity information to the killing chain. The proposed attack detection method necessitates prior knowledge even if it compensates for the lack of high-order attack speech information in the provenance graph. 
MORSE\cite{MORSE} avoids dependency explosion through tag attenuation and tag attenuation technology. 
\vspace{-0.1cm}
\section{Method design}
The framework of our method proposed is shown in Figure \ref{figure:TBDetector frame diagram}. 
TBDetector is mainly divided into three stages: 
In the first stage, TBDetector converts the provenance graph reflecting system changes into a histogram and then extracts it into a feature sequence; 
In the second stage, a sequence feature extractor is trained using a self-attention based encoder-decoder model and then extracts the long-term contextual features of the feature sequence; 
In the third stage, a detection model is built using the anomaly score of each sample, which is determined by the isolation score and similarity score. 
To fully describe the three phases, this section will be divided into three sub-sections. 
\vspace{-0.1cm}
\subsection{Feature Sequence Extraction}
\begin{figure}[h]
  \centering
  \includegraphics[width=220pt]{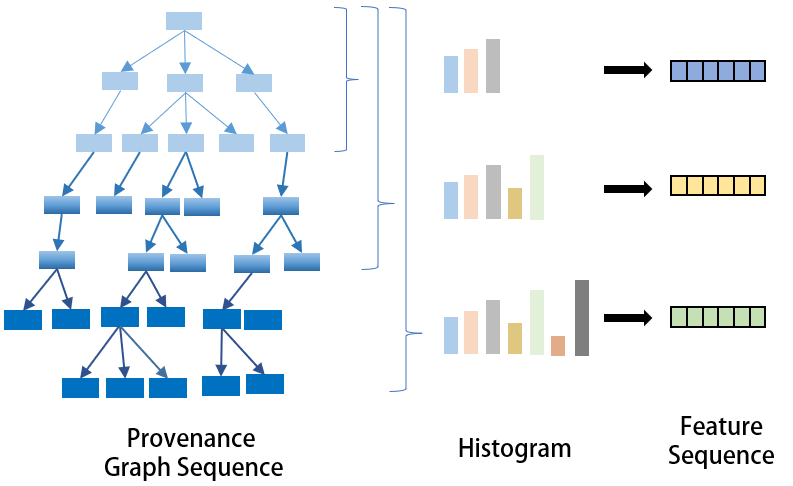}
  \caption{Feature sequence extraction}
  \label{figure:Feature Sequence Extraction}
  \vspace{-0.6cm}
\end{figure}
The frame work of the Feature Sequence Extraction is shown in Figure \ref{figure:Feature Sequence Extraction}. 
In the system framework diagram, the first module's task is to extract the feature sequence of the provenance graph. 
We extract the system call log flow into a feature sequence that can reflect changes in the system state. 
This module implements this operation using the method in UNICORN\cite{Unicorn}, which is primarily separated into three steps: 
In the first step, we obtain the system's call logs to create a DAG provenance graph with a partial order relationship using the Camflow tool. 
The second step is to generate node labels for the provenance graph node and its neighbors' node information and establish a streaming histogram. 
The third step is to convert the histogram into a feature sequence.

The first step is to generate a provenance graph of the system. 
We gather the call log of the system, obtain the subject and object out of the log as the nodes of the provenance graph, 
and then generate the edges of the provenance graph according to the call relationship between the nodes. 
Then use the camflow\cite{CamFlow} tool to build the system call log into a DAG provenance graph with a partial order relationship. 
The constructed provenance graph makes stream computing and context analysis efficient. 

The second step is to convert a provenance graph into a histogram. 
We build a runtime memory streaming histogram that represents the entire history of system execution. 
When new edges enter the graph data stream, we update the count of the histogram's elements. 
By iteratively exploring the larger graph neighborhood, it can discover causal relationships between system entities that provide execution context. 
Considering both the heterogeneous labels on vertices and edges in the substructure and the temporal order of those edges, 
each element in the histogram describes a unique substructure in the graph. 

\begin{figure*}
    \centering
    \includegraphics[width=17cm]{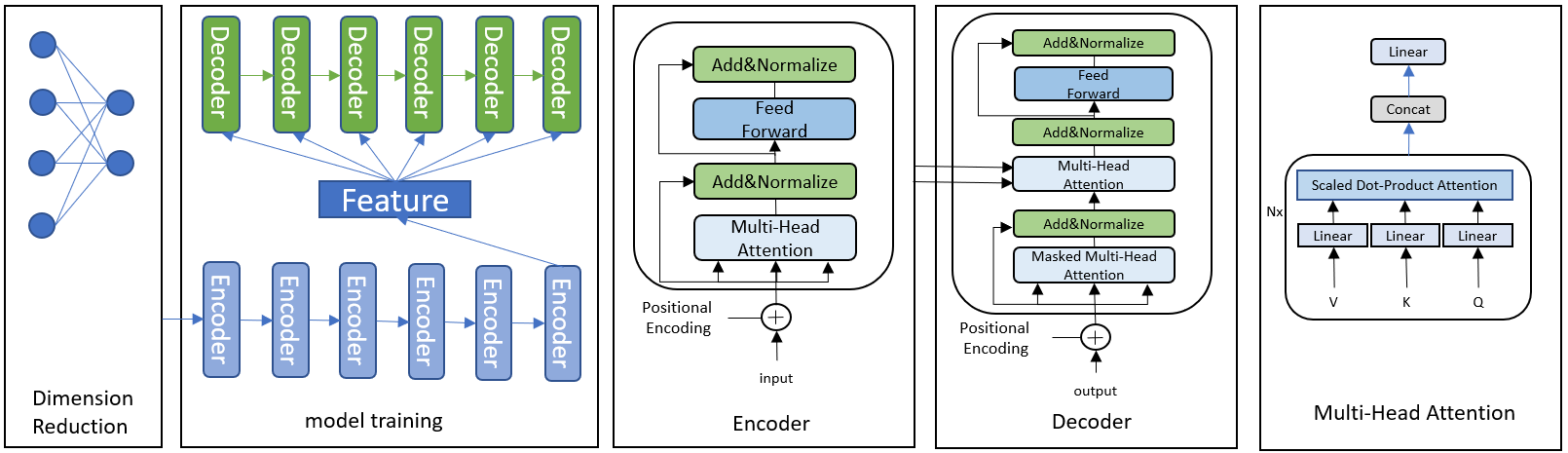}
    \caption{Encoder and decoder based on transformer self attention mechanism}
    \label{figure:Encoder and decoder based on transformer self attention mechanism}
    \vspace{-0.6cm}
  \end{figure*}

The third step is to convert the histogram corresponding to the provenance graph into a feature vector. 
During the process of converting the provenance graph to a histogram, 
node labels that have not appeared may be generated, 
resulting in an indeterminate number of abscissas of the histogram. 
The HistoSketch\cite{HistoSketch} algorithm can effectively solve this problem, 
and successfully convert the histogram into a fixed-length feature vector. 

\vspace{-0.1cm}
\subsection{Long-Term Contextual Features Extraction}
The provenance graph produced by the system call log transformed into a feature sequence in the feature sequence extraction module. 
In this module, the feature sequence is converted into the corresponding feature vector. 
The overall algorithm flow is shown in Figure \ref{figure:Encoder and decoder based on transformer self attention mechanism}. 
Record the feature sequence as $G=\lbrace S_{1},S_{2},S_{3},\dots,S_{n}\rbrace$, where $S_{i}\in \mathbb{R}^d$ represents the i feature vector, 
d is the dimension, and n is the length of the sequence. 
Modeling operations on these sequences are necessary after generating feature sequences that describe system changes. 
Automata and GRU models are used in the earlier methods, but they are unable to fully obtain the sequence's feature information. 
Most of the methods are still based on convolution and loop operations, which led to the increased of detection time. 
As a result, the transformer\cite{attentionTransformer} based on the self-attention mechanism is used in TBDetector. 
Since it is difficult to collect attack data and model all attack behaviors in a real environment, 
it is not suitable to use both normal behaviors and anomaly behaviors to train the model. 
The TBDetector model is trained using just normal data in this method. 
This module is mainly divided into two parts: (1)data preprocessing, (2)model training and sequence feature extraction. 

\textbf{(1)Data preprocessing stage.}
In the input feature sequence $G$, the feature $S_{i}$ in the sequence describes the state of the system at time $i$. 
For two adjacent system states, some of their operations and the corresponding feature vectors are very similar. 
A large amount of redundant data is stored in the feature sequence results. 
A full connection layer needs to be used to reduce the dimensions of all features in the sequence. 
Therefore it is crucial to generate more compact feature vectors in low dimensional space to remove redundant information, and reduce the amount of computation. 
The model adopts the self-attention mechanism and does not include recursion/collaboration, so it cannot capture sequence order information. 
For example, if $K$ and $V$ are scrambled by line, the results are the same after attention. 
However, we must employ the relative or absolute position information of the sequence token since the sequence information is crucial and represents the global structure. 
The input information also needs to be encoded to add the location information of the data. 
We add the original input and position embedding to form the final embedding as the input of the encoder/decoder. 
As the only source of location information in the model, position embedding is a core component of the model and not an attribute of ancillary properties. 
The calculation formula for position embedding is as follows: 
\begin{equation}
  \scriptsize
PE_{(pos,2i)}=\sin(\frac {pos}{10000^{2i/d_{model}}})
\end{equation}
\begin{equation}
  \scriptsize
PE_{(pos,2i+1)}=\cos(\frac {pos}{10000^{2i/d_{model}}}). 
\end{equation}
\textbf{(2)Model training and sequence feature extraction stage. }
TBDetector uses the transformer mechanism of the self-attention based encoder-decoder to train the network by minimizing the reconstruction error. 
This section will introduce self-attention mechanism and encoder-decoder architecture respectively. 
Since the implementation of the encoder and decoder depends on attention, we start with the attention mechanism. 
The attention mechanism is used to calculate the "degree of correlation", and the output is the weight of all values in $V$. 
The weight is calculated by Query and Key. The calculation method is divided into three steps: 

Step 1: Calculate and compare the similarity between $Q$ and $K$, expressed by $f$ as follows 
\begin{equation}
  \scriptsize
f(Q,K_{i}),i=1,2,\dots,m. 
\end{equation}
Step 2: Softmax and normalize the obtained similarity. The formula is as follows 
\begin{equation}
  \scriptsize
a_{i}=\frac {e^f(Q,K_{i})}{\sum^{m}_{j=1}f(Q,K_{j})},i=1,2,\dots,m. 
\end{equation}

Step 3: Determine the attention vector by calculating the weight sum of all values in $V$ for the computed weight, as shown below 
\begin{equation}
  \scriptsize
\sum^{m}_{i=1}a_{i}V_{i}. 
\vspace{-0.3cm}
\end{equation}

Typically, there are four types of calculations method used in the first step: 
dot product  $f(Q,K_{i}=Q^{T}K_{i})$ , general  $f(Q,K_{i})=Q^{T}WK_{i}$ , concat  $f(Q,K_{i})=W[Q;K_{i}]$ , perceptron  $f(Q,K_{i})=V^{T}tanh(WQ+UK_{i})$. 
The attention used in TBDetector is scaled dot-product attention, which is a normalized dot-product attention. 
The input consists of queries and keys of dimension $d_{k}$, and values of dimension $d_{v}$. 
We compute the dot products of the query with all keys, divide each by $\sqrt{d_{k}}$, 
and then apply the softmax function to calculate the weight, as shown below 
\vspace{-0.2cm}
\begin{equation}
  \scriptsize
Attention(Q,K_{i},V_{i})=softmax(\frac {Q^{T}K_{i}}{\sqrt{d_{k}}})V_{i}. 
\vspace{-0.3cm}
\end{equation}

Query, keys, and values are described as matrices Q, K, and V respectively to calculate the output matrix. The formula is shown below 
\vspace{-0.2cm}
\begin{equation}
  \scriptsize
Attention(Q,K,V)=softmax(\frac{Q^{T}K}{\sqrt{d_{k}}})V. 
\vspace{-0.3cm}
\end{equation}
It is not enough to perform such weight operations on Q, K and V only once, therefore multi-head attention is proposed. 
First, do a linear mapping for $Q$, $K$, $V$, and map the $Q$, $K$, $V$ matrices whose input dimensions are $d_{model}$ to $Q\in R^{m\times d_{k}}$, $K\in R^{m\times d_{k}}$, $V\in R^{m\times d_{v}}$. 
Then calculate the result with scaled dot product attention. 
We iterate the above two steps several times and merge the results of $V$. 
Then we perform the linear transformation on the combined results, as shown below 
\begin{equation}
  \scriptsize
Attention(Q,K,V)=Concat(head_{1},head_{2},\dots head_{h})W^{(0)}
\end{equation}
\begin{equation}
  \scriptsize
head_{i}=Attention(QW_{i}^{Q},KW_{i}^{K},VW_{i}^{v}). 
\vspace{-0.3cm}
\end{equation}

The encoder consists of six layers, each with two sub-layers. 
The first sub-layer is the multi-head self-attention layer, which is used to calculate the input self-attention. 
The second sub-layer is a simple fully connected network. 
In each sub-layer, we simulate the residual network and the output of each sub-layer is shown below 
\vspace{-0.1cm}
\begin{equation}
  \scriptsize
LayerNorm(x+Sublayer(x)). 
\vspace{-0.1cm}
\end{equation}
$Sublayer(x)$ refers to the mapping of input $x$ by sub-layer. 
The decoder consists of six layers, each with three sub-layers. 
To ensure connectivity, we set all sub-layers and embedded layers to have the same size. 
The first sub-layer is masked multi-head self-attention. 
The value in front of $i$ is known when we anticipate the $ith$ value, but the value behind $i$ is not calculated, so mask operation is necessary. 
The second sub-layer is a fully connected network which is the same as the encoder. 
The third sub-layer is to calculate the attention of the encoder input. 
At the same time, the self-attention layer in the decoder needs to be modified, 
because only the input before the current time can be obtained. 
Then the input before time $t$ is used for the attention calculation, 
which is also called the mask operation. 

There are three multi-head attention modules in the encoder-decoder architecture. 
The first is the self-attention of the encoder module. 
In encoder, the self-attention input $Q,K,V$ of each layer is the output of the previous layer and $Q=K=V$. 
Each position in the encoder can obtain the output of all positions in the previous layer. 
The second is the mask self-attention of the decoder module. 
In the decoder, each position can only obtain the information of the previous position, so you need to mask and set it to $-\infty$. 
The third is the attention between the encoder and the decoder, 
where $Q$ comes from the output of the previous decoder layer and $K, V$ come from the encoder's output. 
Allowing each position of the encoder can obtain all the position information of the input sequence. 
For position-wise feed forward networks, after the attention operation, each layer in the encoder and decoder contains a fully connected forward network. 
The same operations are performed on the vectors of each position, including two linear transformations and a RELU activation output. 
The parameters of each layer are different. 
The formula is as follows 
\begin{equation}
  \scriptsize
FFFN(x)=max(0,xW_{1}+b_{1})W_{2}+b_{2}. 
\vspace{-0.3cm}
\end{equation}

The self-attention based encoder-decoder are used to train the feature extractor by reconstructing the minimum error of the input sequence. 
Then the feature extractor is used to obtain the feature vector of the sequence. 

\subsection{Anomaly Detection}
In real-world applications, it is easier to obtain normal behavior data, while it is more difficult to obtain anomaly behavior data. 
So in order to identify the unknown anomaly more effectively we use the normal behavior to construct the detector. 
We use the similarity score and the isolation score to calculate the anomaly score for each sample, and build a detection model to detect APT attacks. 
This stage follows a two-stage approach. 

In the first stage, the observed normal and unlabeled samples are processed. 
First, we use K-means to cluster the observed normal data to generate $K$ clusters $C={C_{1},C_{2},\dots,C_{k}}$. 
The distance between two samples is measured using squared euclidean distance, as shown below 
\begin{equation}
  \scriptsize
dist(x_{i},x_{i}^{'})=\sum^{d}_{j=1}(x_{ij}-x_{i^{'}j})^{2}. 
\end{equation}
Here $d$ is the dimension of the samples. 
The similarity score is calculated based on the distance between the sample and the cluster center. The formula is as follows 
\begin{equation}
  \scriptsize
SS(x)=\max_{i=1}^{k}e^{-(x-\mu_{i})^2}. 
\end{equation}

The concept of isolation forest is introduced to calculate the isolation score. 
Each tree in the forest is constructed by randomly choosing an attribute and corresponding split value for subsequent growth at each node. 
Anomaly samples will always be isolated closer to the root of the tree, while normal samples will go into the deeper leaves of the tree\cite{ADOA}. 
Based on the average path lengths on the trees, the isolation score $IS(x)$ can be calculated to describe the probability of a sample $x$ being an anomaly. 
Let $h(x)$ represent the path length of a sample $x$ on a tree, and $E(h(x))$ represent the average path length of a collection of isolation trees. 
The symbol $n$ represents the sample size. 
The H(n) is the harmonic number, which can be estimated by $ln(n) + 0.5772156649$ (Euler's constant). 
The formula is as follows 
\vspace{-0.1cm}
\begin{equation}
  \scriptsize
IS(x)=2^{\frac {-E(h(x))}{(2H(n)-(2(n-1)/n))}}. 
\vspace{-0.3cm}
\end{equation}

In the second stage, 
we set $\alpha$ to indicate the average score of observed anomalies, as follows 
\vspace{-0.1cm}
\begin{equation}
  \scriptsize
\alpha=\frac{1}{L}\sum_{i=1}^{L}TS(x_{i}). 
\vspace{-0.2cm}
\end{equation}
The total score is calculated by the isolation score and the similarity score. 
If $TS(x)>\alpha,$ the sample $x$ is divided into a potentially normal. 
If $TS(x)<\alpha,$ the sample $x$ is divided into anomaly, as follows 
\vspace{-0.1cm}
\begin{equation}
  \scriptsize
TS(x)=(1-\theta)SS(x)+\theta IS(x), 
\vspace{-0.2cm}
\end{equation}
in which $\theta \in [-1, 0]$ is a parameter to balance the weighting of isolation score and similarity score. 
The value of $\theta$ we determined through testing. 

We set the anomaly score of all known normal points to 1. 
The anomaly score of the unlabeled sample can be calculated by the following formula, as shown below 
\vspace{-0.1cm}
\begin{equation}
  \scriptsize
AS(x)=\frac {TS(x)}{\max TS(x)}. 
\vspace{-0.2cm}
\end{equation}
Then we combine the anomaly score with the classifier to determine whether it is abnormal data. 
\vspace{-0.1cm}
\section{Experiment}
\subsection{The Testing Datasets}
\vspace{-0.2cm}
\begin{table}[!htbp]
  \scriptsize
  \begin{center}
  \caption{Characteristics of the StreamSpot dataset}
  \begin{tabularx}{28em}
  {*{4}{>{\centering\arraybackslash}X}}
      \toprule
      Dataset  & Label & Graphs & Data Size(GiB)\\
      \midrule
      \multicolumn{1}{c}{\multirow{6}*{StreamSpot}} & YouTube & 100 & 0.3\\
		  \cline{2-4} & Gmail & 100 & 0.1\\
      \cline{2-4} & Download & 100 & 1\\
      \cline{2-4} & VGame & 100 & 0.4\\
      \cline{2-4} & CNN & 100 & 0.9\\
      \cline{2-4} & Attack & 100 & 0.1\\
      \bottomrule
  \end{tabularx}\label{table:Characteristics of the StreamSpot dataset}
  \end{center}
  \vspace{-0.5cm}
\end{table}

\begin{table}[!htbp]
  \scriptsize
  \begin{center}
  \caption{Characteristics of the DARPA TC dataset}
  \begin{tabularx}{28em}
  {*{4}{>{\centering\arraybackslash}X}}
      \toprule
      Dataset  & Label & Graphs & Data Size(GiB)\\
      \midrule
      \multicolumn{1}{c}{\multirow{2}*{ClearScope}} & Benign & 43 & 441\\
		  \cline{2-4} & Attack & 51 & 432\\
      \multicolumn{1}{c}{\multirow{2}*{CADETS}} & Benign & 110 & 271\\
      \cline{2-4} & Attack & 3 & 38\\
      \bottomrule
  \end{tabularx}\label{table:Characteristics of the DARPA TC dataset}
  \end{center}
  \vspace{-0.5cm}
\end{table}

\begin{table}[!htbp]
  \scriptsize
  \begin{center}
  \caption{Characteristics of the SC dataset}
  \begin{tabularx}{28em}
  {*{4}{>{\centering\arraybackslash}X}}
      \toprule
      Dataset  & Label & Graphs & Data Size(GiB)\\
      \midrule
      \multicolumn{1}{c}{\multirow{2}*{wget}} & Benign & 125 & 64\\
		  \cline{2-4} & Attack & 24 & 12\\
      \multicolumn{1}{c}{\multirow{2}*{shellshock}} & Benign & 125 & 59\\
      \cline{2-4} & Attack & 25 & 12\\
      \bottomrule
  \end{tabularx}\label{table:Characteristics of the SC dataset}
  \end{center}
  \vspace{-0.5cm}
\end{table}

In our experiments, we employ five widely used anomaly detection datasets to evaluate the performance of our method, \ie, 
StreamSpot dataset, ClearScope dataset, CADETS dataset, wget dataset and shellshock dataset. 
More specifically, the StreamSpot dataset contains a total of 6 scenarios, as shown in Table \ref{table:Characteristics of the StreamSpot dataset}. 
It contains five normal scenarios and one anomaly scenario. 
The attack scenario is to download a program from a malicious url and take advantage of flash memory vulnerability to gain system administrator privileges. 
Each scenario was run 100 times. 
The DARPA TC dataset is from the third two-week offensive-defense confrontation, as shown in Table \ref{table:Characteristics of the DARPA TC dataset}. 
Clearscope is collected on the android system. 
The SC dataset is collected by the authors of UNICORN. 
A total of two APT attack scenarios wget and shellshock are simulated in the SC dataset as shown in Table \ref{table:Characteristics of the SC dataset}. 

\subsection{Evaluation Criteria}
Precision and recall are two crucial evaluation criteria utilized in detection issues. 
Precision is a measure of the classifier's ability to correctly identify samples, 
and it represents the proportion of correctly predicted samples among the samples identified as positive samples. 
The recall represents the proportion of all positive samples that can be predicted correctly. 
The calculation formulas are as follows: 
\begin{figure*}
  \centering
  \includegraphics[width=17cm]{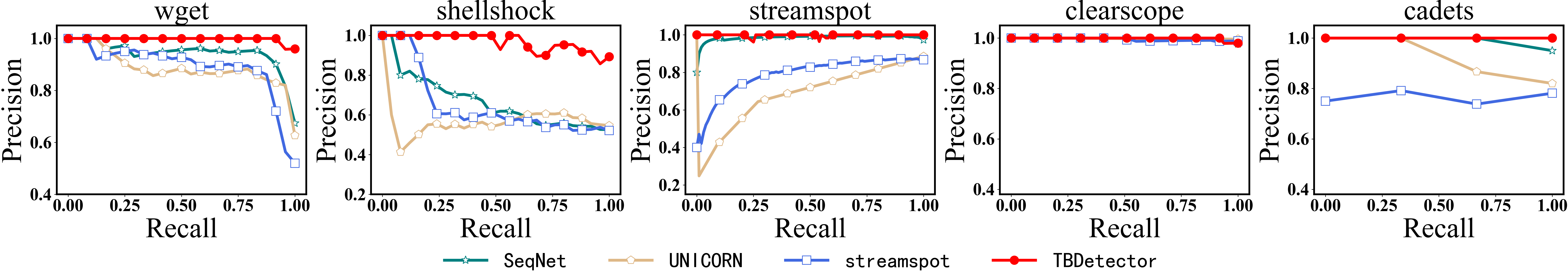}
  \caption{Experimental results}
  \label{figure:Experimental results of four methods}
  \vspace{-0.3cm}
\end{figure*}
\vspace{-0.2cm}
\begin{equation}
  \scriptsize
precision=\frac {tp}{tp+fp}
\end{equation}
\begin{equation}
  \scriptsize
recall=\frac {tp}{tp+fn}.
\vspace{-0.2cm}
\end{equation}
The method proposed in this paper is to model normal behavior for attack detection. So thresholds need to be set during the testing phase. 
However, there is no fixed optimal threshold in all real environments, 
we draw the PR (precision-recall) curves by obtaining the precision and recall under various thresholds. 
We use the area of the PR curve as the evaluation Criteria. 
The curve area of recall and precision is directly proportional to the superiority of the method. 
\subsection{Compared Methods}
\begin{figure*}
  \centering
  \includegraphics[width=17cm]{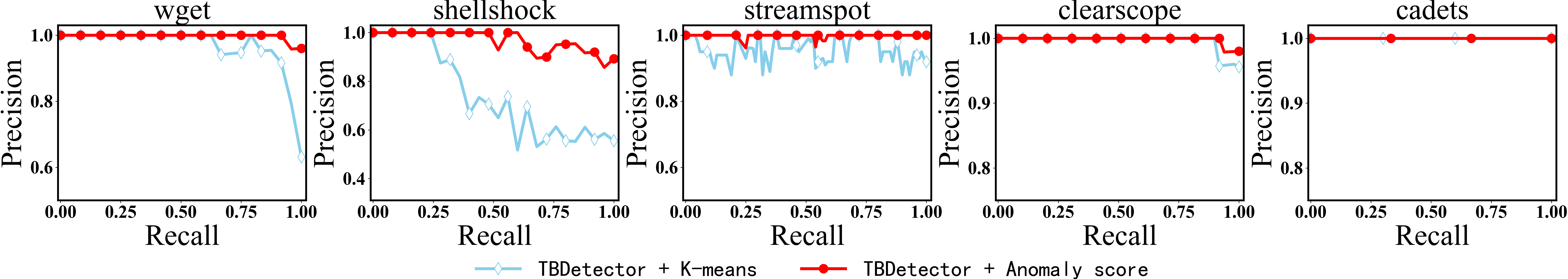}
  \caption{Verify anomaly score's effectiveness}
  \label{figure:Comparison between TBDetector and K-means}
  \vspace{-0.2cm}
\end{figure*}

To evaluate the effectiveness of the proposed algorithm, 
we compare our proposed TBDetector method with three state-of-the-art methods in order to assess the efficacy of the proposed algorithm, 
and all of the parameters of the compared method are set as the proper values. 
The lines in each figure contain many points but only a few are labeled for the sake of the graph's aesthetics. 
Three of the most state-of-the-art methods are listed below: 
\vspace{-0.2cm}
\begin{enumerate}[(1)]
\item streamspot\cite{streamspot} deals with streaming heterogeneous graphs. It is an anomaly detection system based on clustering. Streamspot analyzes streaming information graphs to adhere to abnormal activities, but the graph features are subject to local constraints. 
\item UNICORN\cite{Unicorn}  is an anomaly-based APT detector and can effectively use the data Provenance analysis to realize the end-to-end anomaly attack detection model. 
\item SeqNet\cite{seqnet} realizes feature extraction of sequences by GRU combined with local attention mechanism, and further realizes detection of abnormal data by Kmeans. 
\end{enumerate}
\vspace{-0.5cm}

\subsection{Experimental Results}
Compared with state-of-the-art methods, \ie, StreamSpot, UNICORN, SeqNet outperforms the compared method on five datasets consistently. 
The experimental results are shown in Figure \ref{figure:Experimental results of four methods}. 

More specifically, when 20\% of data are used for training, in terms of shellshock, 
the PR area corresponding to TBDetector is 0.9575, an increase of 46.83\% compared to SeqNet's 0.6521, 
an increase of 75.43\% compared to unicorn's 0.5458, and an increase of 50.87\% compared to streamspot's 0.6347. 
When 10\% of data are used for training, in terms of cadets, 
the PR area corresponding to TBDetector is 1.0, an increase of 1.69\% compared with 0.9833 of SeqNet, 11.66\% compared with 0.8956 of unicorn, 
and 34.76\% compared with 0.7421 of streamspot. 
When 25\% of data are used for training, in terms of streamspot, 
the PR area corresponding to TBDetector is 0.9967, an increase of 1.24\% compared with 0.9844 of SeqNet, 46.66\% compared with 0.6796 of unicorn, 
and 26.52\% compared with 0.7878 of streamspot. 

Compared with the UNICORN algorithm, streamspot algorithm and SeqNet algorithm, 
the method proposed in this paper achieves better overall results, 
mainly because the TBDecetor model can capture long-term contextual features. 
In real network attack scenarios, the APT attack cycle and corresponding feature sequence are quite lengthy. 
The method proposed in this paper can extract long-term context patterns more effectively, 
making it better able to adapt to the characteristics of APT attacks in real scenes and produce better detection results. 
Compared with the streamspot algorithm, The advantage of this paper is that the provenance graph sequence is extracted into a sequence of eigenvectors. 
The eigenvectors can effectively reflect the state of the system, and the sequence of eigenvectors contains the time-series features of system state changes. 
The time-series features of this sequence can be efficiently utilized using the method in this study. 
Although the UNICORN algorithm also uses the feature sequence, it matches the state machine template. 
Anomaly activity will only be recognized as such when there is a significant difference between it and normal behavior; otherwise, it will be ignored. 
The original convolution and cyclic forms used by the SeqNet algorithm for data processing cannot fully guarantee timeliness. 
In the process of capturing sequence features, the transformer model is used to fuse the historical and current features of the sequence. 
The calculated sequence features combine the characteristics of the entire sequence and can keep track of and collect the sequence's tiny variations. 
At the same time, 
the model can now capture sequence information more quickly and precisely thanks to the introduction of the self-attention mechanism, 
which replaces the old calculation modes of circulation and convolution throughout the model training process. 
\begin{figure*}
  \centering
  \includegraphics[width=17cm]{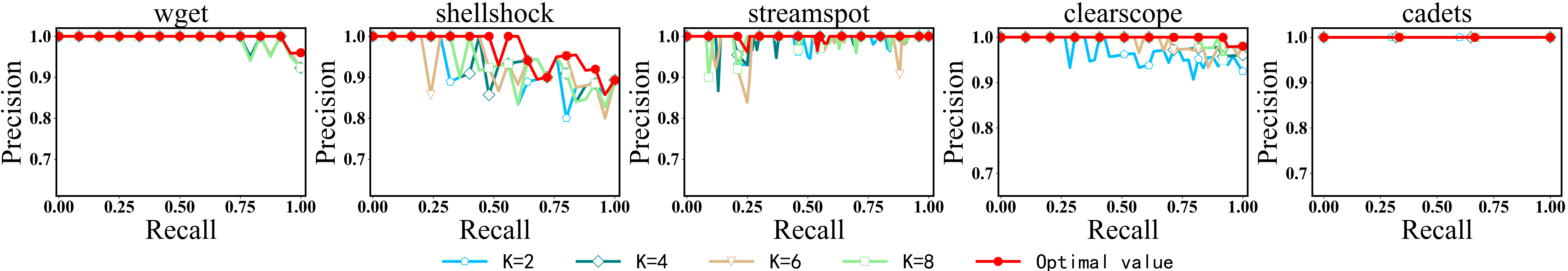}
  \caption{Influence of parameter K on experimental results}
  \label{figure:Influence of parameter K on experimental results}
  \vspace{-0.5cm}
\end{figure*}
\subsection{Verify Anomaly Score's Effectiveness}
After obtaining the feature vector of the feature sequence, TBDetector employs anomaly scores to detect the APT attacks. 
Anomaly score is calculated by similarity score and isolation score. 
We build a detection model to detect APT attacks. 
In both UNICRORN and SeqNet, K-means is added at the end of the method to realize attack detection, 
so we compare the anomaly score with Kmeans to show the advantages of our method. 
This detection method is more effective for the more popular K-means. 
The contrast effect is shown in Figure \ref{figure:Comparison between TBDetector and K-means}. 

More specifically, when 20\% of data are used for training, in terms of shellshock, 
the PR area corresponding to TBDetector is 0.9575, an increase of 33.69\% compared to K-means's 0.7162. 
When 20\% of data are used for training, in terms of wget, the PR area corresponding to TBDetector is 0.9965, 
an increase of 3.89\% compared to K-means's 0.9592. 
when 25\% of data are used for training, in terms of streamspot, 
the PR area corresponding to TBDetector is 0.9967, an increase of 5.54\% compared to K-means's 0.4645. 
We can see that the anomaly score used in this paper has the same or better effect than the previous K-means. 

\subsection{Parameter Experiment}
The anomaly score is calculated from the similarity score and the isolation score. 
In the process of calculating the similarity score, it is necessary to select K cluster centers for some known normal samples. 
The situation in real environment is complex and changeable, and the number of cluster centers for data is not fixed. 
To evaluate the influence of clustering number K on the experimental results. 
In this section, we set the values of K to $\{2, 4, 6, 8\}$ on the five data sets, respectively, 
to show the PR curve area corresponding to each K value, and compare it with the use of anomaly score. 

The experimental results are shown in Figure \ref{figure:Influence of parameter K on experimental results}. 
It can be seen from the figure that with the increase of the K value, the precision value of the same recall value does not show a fixed upward trend. 
Overall, the optimal values produced are not completely concentrated in one cluster number if the number of clusters is fixed. 

More specifically, when 20\% of data are used for training, in terms of shellshock, 
the PR area corresponding to TBDetector is 0.9575, 
an increase of 5.63\% compared to K=2's 0.9065, 
an increase of 3.93\% compared to K=4's 0.9213, 
an increase of 4.97\% compared to K=6's 0.9122, 
and an increase of 4.15\% compared to K=8's 0.9193. 
When 10\% of data are used for training, in terms of clearscope, 
the PR area corresponding to TBDetector is 0.9983, 
an increase of 3.74\% compared to K=2's 0.9623, 
an increase of 0.78\% compared to K=4's 0.9906, 
an increase of 1.16\% compared to K=6's 0.9869, 
and an increase of 0.37\% compared to K=8's 0.9946. 
Therefore, TBDetector generates a different number of cluster centers for different test samples, which is more effective than a fixed K value. 
The experimental results show that the range traversal optimal number of clusters used by TBDetector is generally more efficient. 
\section{CONCLUSION}
We propose TBDetector, a real-time anomaly detection. 
TBDetector generates the log information into a provenance graph and converts it into long sequence features. 
Then the encoder-decoder model of self-attention mechanism is used to train a sequence feature extractor, 
which can effectively extract long-term contextual features. 
The obtained contextual features make it easier to distinguish normal data from anomaly data. 
The similarity score and isolation score are combined to calculate the anomaly score for each sample, and a detection model is created to detect APT attacks. 
Experimental results and comparisons with the advanced methods have exhibited better performance of our proposed method.


\bibliography{example_paper}
\bibliographystyle{icml2023}


\end{document}